**International Conference on New Interfaces for Musical Expression**

# The Body Electric: A NIME designed through and with the somatic experience of singing

**Kelsey Cotton[1], Pedro Sanches[1], Vasiliki Tsaknaki[2], Pavel Karpashevich[1]**

**[1]KTH Royal Institute of Technology, [2]Digital Design, IT University of Copenhagen**

**ABSTRACT**

This paper presents the soma design process of creating *Body Electric:* a novel interface for the capture and use of biofeedback signals and physiological changes generated in the body by breathing, during singing. This NIME design is grounded in the performer's experience of, and relationship to, their body and their voice. We show that NIME design using principles from soma design can offer creative opportunities in developing novel sensing mechanisms, which can in turn inform composition and further elicit curious engagements between performer and artefact, disrupting notions of performer-led control. As contributions, this work 1) offers an example of NIME design for situated living, feeling, performing bodies, and 2) presents the rich potential of soma design as a path for designing in this context.

# 1. Introduction

The development and commercial availability of biofeedback sensors, such as electrocardiograms (ECG) and electromyograms (EMG), has seen the increased usage of these technologies within an artistic context over the last 60 years [1]. The creative appropriation of biosensors has resulted in the development of new musical interfaces which harness biosignals, spatial and gestural data to either sonify the body, control and trigger sound events, or to manipulate the sound of acoustic instruments. Some prominent artists working with these technologies such as Marco Donnarumma, Atau Tanaka, Pamela Z and Imogen Heap, have approached the design of their NIMEs and digital instruments for sonifying their bodies through harnessing electrical signals from muscles and/or the brain—in Donnarumma and Tanaka's work— [2] [3] [4] [5] [6] [7], or—as with Z and Heap— through sonifying  gestural data [8] [9][10] [11] [12] [13] [14] [15] [16].

However, there appears to be a gap in the design of interfaces for controlling and processing vocal sound. Although there is notable ongoing work and research in voice and the application of biofeedback and biosensing technologies, a significant majority is geared towards discovering how the expressivity of a singer's gestures (typically their hands) can be harnessed to process their sound according to sonification mappings based on spatial orientation and acceleration of limbs [17][18], or the recognition of a gestural vocabulary [19][14][8]. Recent research and applications of respiration as a control parameter within NIMEs have seen the development of novel interfaces and instruments, and investigated couplings between music control, biofeedback and breathing [20], or a dancer's respiratory influence upon music [21]. These works have considered features of respiration, such as air pressure in the mouth of wind instrument players [22], periodicity of the breathing cycle and the coupling between affective health [20], frequency content of the air stream during extended and timbral techniques whilst playing wind instruments [23], and the sonification of breathing data within interactive systems during free improvisation [24]. Of most connection to this research is Lee and Yeo's [21] observations on dancers' respiratory connection to the motion of the body, through torso movements. Building upon this connection, this paper explores topographical change in the torso occurring during singing, breathing and breathing for singing. At the time of writing, there is a dearth of research examining how the physiological processes associated with producing vocalised sound - such as respiration - can be harnessed as a means of interacting with and processing the voice itself during singing.

Here we present a novel approach to interface design by "turning inwards" to the rich bodily information provided during singing, framing the design and creation of a NIME within the experience of (or relationship to) the instrument being sensed and designed for. By "turning inwards", we define this as a utilisation of a first-person subjective somatic experience as a method to articulating their experience (in this case, singing), and further, how the experiential qualities of vocal performance are designed for when developing a musical interface.

The first author, Kelsey, is a classically trained vocalist concerned with the actions occurring within her body during performance, and wished to highlight the experiential and physiological processes involved in her singing. She wanted to compose and perform a piece of music around her interest, and to create an interface that emphasised and interacted with the invisible aspects of her vocal production. Throughout this paper we will use the term "instrument" to refer to Kelsey's voice, with more general usage of the term framed within the context of usage (ie. other musical instruments, digital music instruments) with the terms 'interface' or 'artefact' referring to the *Body Electric* NIME.

Kelsey's intention—to place the audience "inside" her bodily experience of singing, and thereby draw attention to how she "feels" her instrument and the invisibility of the physiological changes within her body (through the intake and flow of her breath).Motivated by this, an interface, in the form of an interactive wearable utilising novel shape change pressure sensor-actuators to capture breathing biosignal information, was designed. This became a sort of co-performer in the resultant composition: *The Body Electric*, created with and through this interface. To design this, we drew from soma design strategies aimed at engaging with the felt experience of breathing and singing. In soma-based design work, it is important to train somatic designerly skills [25] by shaping not only the physical materials used to build our artefacts, but also our own movements and sensations as we learn to aesthetically appreciate and to articulate the somatic changes and movements occurring when singing [26] [27].

This paper documents this autobiographical engagement with Kelsey's instrument through and with this interface and the resulting creation of a wearable artefact to record, capture and transmit these biosignals. It further documents the evolving relationship and curious entanglements between musician and interface within the context of the composed work (*The Body Electric*) and builds upon an area of interface design centred in the musician's subjective and somatic experience with and through their instrument. We discuss the lessons learned when designing musical interfaces aimed at co-performing with a singer, as well as challenges and opportunities for NIME design when turning inwards, facilitated through soma design methods.

## 2. Background

### Biosensing Technologies for Sonifying the Body in Artistic Contexts.

The intersection of biosensing and art has been a progressively developing field as the commercial availability of sensors has increased, and so too the rise of DIY/Maker culture [1] [28] [29]. Within the NIME community a number of researchers have explored the potential of utilising biosensing tools for generating music [30] or for designing creative and artistic expressions with data produced by bodies, more broadly [31] [32] [33]. Regardless of the type of biosensor used to capture and transform physiological data into sound, there is a commonality in that each approach is centred around exploring how a musician's experiences can inform physical engagement during performance, and the body's creative potential as an expressive tool [34] [11].

Donnarumma's description of biophysical music is useful here, which he defines as the configuration of computing interfaces with musical systems [35]. We see this definition as more effectively accommodating the integration of biosensing and sonification, while opening up this design space through providing opportunities to develop new technologies and means to record and capture novel human movement during performance and music making. Our work takes this definition as a point of departure, and explores the torso as an "unconventional" computing interface that provides information through the control of Kelsey's air pressure system, as it changes shape, topography, dimension and hardness when she breathes and engages with certain muscular regions. Our research process and resulting interface also builds on previous work in NIME that has used somaesthetics and soma design methods to open up the space of developing musical interfaces [36].

With regards to singing and vocal production within the context of NIME and HCI research, there has been notable work [17] [37] [38] [39] [18]. Two who have been quite influential to our research, are Pamela Z and Imogen Heap. Both utilise the gestural components of their respective performative practices as a means of controlling and processing their sound, and as a compositional tool. The vocalist-composer Pamela Z utilises gestural MIDI controllers, and the data from her hand movements is captured and processed using MaxMSP and Isadora software. Singer-songwriter-engineer Imogen Heap utilises sensors monitoring her body and gestures through her Mi-Mu gloves, which capture movement and gestural information using flex, motion and orientation sensors, and haptic motors [14]. The data captured through Heap's Mi-Mu gloves is processed using Ableton Software.

## 3. Motivation and Vision

Through her experience in singing with various interfaces, controllers and sensors (such as Wii remotes, MIDI keyboards, pedals and motion tracking), Kelsey identified that these systems requiring gestural activation—mainly through the hands) or extraneous movement for motion detection—"pushed" the manipulation of her sound outside of her body, disconnecting her from her own practice in the embodied manipulation of her own acoustic sound by engaging with her own musculature. This was experienced as a de-centralisation of her body as an interface in its own right, and the physiological and experiential movements and actions of Kelsey generating a sonic action - in this case, the changing topography of her torso and abdominals during singing and breathing. To address this potential, a NIME was developed to re-position the body as the controller, or interface, for the sound coming out of it and to enable reflection on the somatic experience of singing through topographical change to the torso.

## 4. Soma Design Approach: Developing a NIME through and with the Body

Aiming to achieve the above goals (i.e. design a NIME that would put her bodily manipulations at the centre of the experience and explore how to "place" the audience inside her body by emphasising the less visible bodily changes she makes during singing and breathing) Kelsey turned to soma design, as a method and approach in interaction design that puts the focus on designing with and for the body [40]. Soma design has its roots in theories of somaesthetics [41] and emphasises becoming attentive to, and improving connections between movement, sensation, subjective understanding and values. It is building on the designer's first person perspective of aesthetic and sensuous experiences, through which, one can learn to discern and harness them, and to devise rich and nuanced interactive experiences at the meeting between technologies and bodies. The most common soma design tactics include becoming attentive to one's soma through engaging in bodily practices thoughtfully and deliberately, and using estrangement methods [42] for defamiliarising already familiar experiences as a path to providing a more nuanced and rich perception of bodily experiences. Martinez-Avila et al. [36] were the first to turn to soma design methods as a potential route to realising more body-centred and aesthetic musical instruments and interfaces. With their concept of *breathing guitars,* arising through their engagement with soma design for NIMEs, they suggest that there is great potential for further applying the soma design approach within the NIME community. In this paper we draw on this approach and open up the design space of making NIMEs that have the body, and the subjective somatic experience of the body, as the starting point of departure and as the actual musical interface. We show a process of designing the *Body Electric* NIME—a musical interface that firmly positions the body and the experience of breathing while singing—into the heart of the experience of performing with it.

## 5. THE DESIGN PROCESS

### 5.1 Somatic Experience of Singing

The first step in this design process was turning inwards to Kelsey's experience of singing and examining closely what she observed her body doing during her vocal practice, she observed the following breathing phenomena occurring: expansion, multi-directional movement and collapse of the torso and thorax. Specifically in the following regions: lower abdominals, ribcage, sternum, thoracolumbar fascia (see Figure 1). These observations came through Kelsey's usage of body mapping to track physiological changes and become more attentive to her body, disrupting her normal breathing exercises by working with different speeds and opposing her body's instinctive movements (ie. pulling in during inhalation, and pushing out for exhalation), and attuning to experience of breathing. Kelsey's observations about the parts of her body that engage deeply in her singing practice (illustrated below) and affected by her topographical manipulation guided the need for sensors capable of registering her body changing shape during her breathing, and where these sensors should be placed.

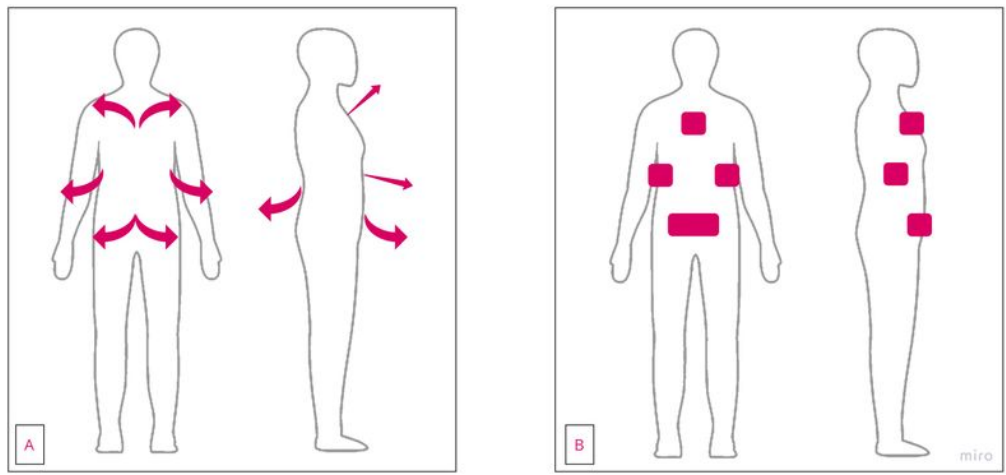

**Figure 1. Body maps detailing breathing and sensor placement.**
Section A: Body map detailing breathing zones on the body. Section B: Placement of shape change pressure sensors according to breathing zones.

During Kelsey's attunement to her body she further observed nuanced experiences of how air feels within her body, such as sonic differences in inhaling through her mouth versus her nose, textural sounds when she breathes deeply or shallowly, and high or low into her body, gurgling of the digestive system as the organs are displaced by inhalation, the tactile crackle of saliva in the mouth during singing and the different filtering of the voice caused by inner-ear pressure and blocked sinuses. These observations influenced Kelsey's use of sonic material, both as live vocalisation and as pre-recorded material.

## 5.2 Harnessing somatic experience: biosensing sensor-actuator shape change pillows

Building upon Kelsey's somatic understanding of her breathing practice (section 5.1), she created a corset structure fitted with 4 pressure sensing pillows for detecting topographical bodily changes across 4 points on the body to capture the physiological changes she experienced. The internal sensor system was developed in collaboration with the co-authors of this paper, who are HCI researchers working at the KTH Royal Institute of Technology, and builds on earlier research and artefact design conducted in developing novel shape-changing sensors [43].

These biosensors took the form of shape change sensor-actuator pillows, specifically designed to capture the shape-changing qualities of Kelsey's own body. A single sensor-actuator consists of an electronics unit with an Arduino micro controller, solenoid valves, an air pump and a barometric pressure sensor (see Figure 2 below). The pillow —made of Thermoplastic Polyurethane (TPU) coated nylon—connects to the air pump with transparent tubing. The Arduino micro controller is connected to a computer via OSC protocol, with software modules in MaxMSP for controlling the inflation and deflation operations of each micro controller. The pillows are coded to respond to external crush by deflating, and—after deflating—to slightly re-inflate when not placed under any external pressure.

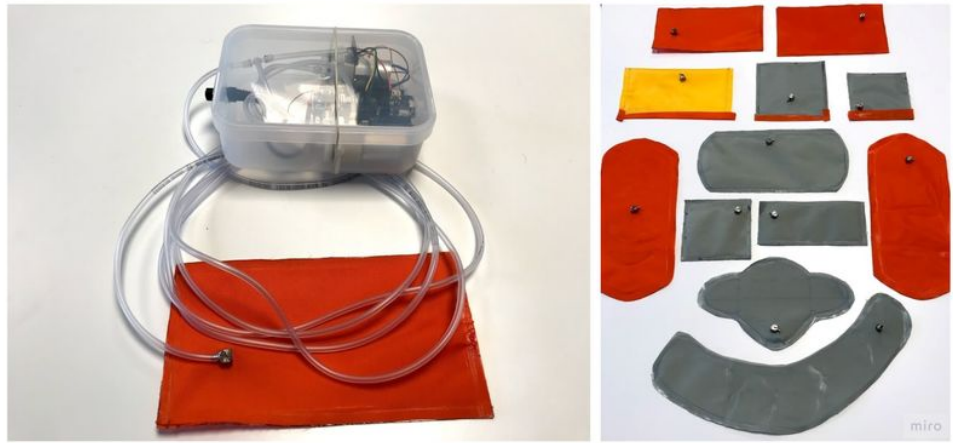

**Figure 2. Shape change sensor-actuators, and pillow shapes**

The pillows were positioned over specific areas (see Figure 1), between Kelsey's body and an external rigid surface. Much like the physical limits imposed by ribcage flexion, an external surface to press against the pillows was needed to capture the torso's changing dimension and shape. To facilitate this, Kelsey constructed a hard external shell which was fitted with the internal pillow sensors (see Figure 3, panels A and B). Within the final version of this NIME (Figure 3, panel C) the movements of the body and "crush" of the 4 sensors are visible, with changes to Kelsey's torso topography captured via the pressure sensors in the pillows.

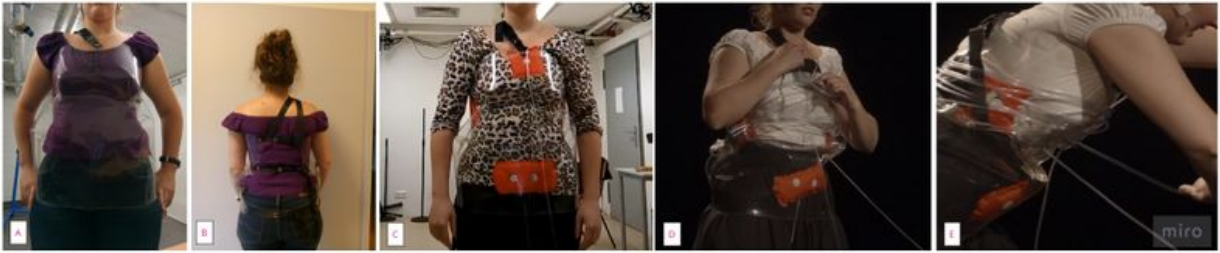

**Figure 3. Completed Body Electric NIME**
Panel A, B: External containment unit. Panel C: External unit and internal sensor system. Panel D, E: The artefact in action

In *The Body Electric*, the corset functions as an interactive interface which utilises biosensing information caused by torso changes (captured through 4 shape change sensor-actuator pillows) occurring as a result of singing and breathing. The use of these sensor-actuators facilitated a novel experience of engaging with the interface, and further re-frames the act of singing as a conscious and deliberate engagement with the physiological events occurring within the body through an interface that continually creates a control versus co-performance environment.

## 5.3 Somatic Singing

### 5.3.1 Body Sounds

The specific sound world built in *The Body Electric* revolves around the internal sounds of Kelsey's instrument as she hears her voice and breath (section 5.1). The interactive nature of singing *in* and *with* the interface further manipulated her sound because she had to "push" out and "hold" her body out more to crush the pillows during singing, inviting deeper interoceptive reflection around her breathing. This approach led to her elongating the sounds she normally hears during singing, such as the micro sounds of air moving in her nasal passages and mouth, the crackle of her saliva and the slower displacement and gurgling of organs in her digestive system displaced slightly by her breathing. Kelsey used this sonic material to try and "place" the audience inside her experience of her air, using these body sounds in a soundscape affected by the micro movements of her breathing and muscles. These sounds were actual sounds produced live (with amplification), and also as material that was recorded for playback and transformation in performance.

### 5.3.2 Composing with the body, composing with the corset: Through connection, disconnection and questioning

The interactive experience of singing *in* and *with* the interface further influenced the composition. During exploratory sessions wearing the corset, Kelsey examined the MaxMSP visualisation of her breathing (from the 4 pressure sensors, see Figure 4) and the specific architecture of each biosignal line generated from the pillow sensors. She observed during these sessions that there was an interesting correlation between the visualisation of the signals and her experience of interaction with the corset: over time the physical fatigue she experiences during singing accelerated due to the interactive nature of singing inside the artefact, where Kelsey would "push" more during inhalation and progressively collapse faster. The "giving out" of her body further led to her breathing pattern became progressively more disrupted as her breath progressively moved higher into her body, and became more fragmented as she took in more air to try and help her "failing" body.  The pillows on the points of her body more effected by fatigue (lower abdominals) became less integral to the soundscape when they "couldn't sing anymore". Kelsey drew connections between the visualisation of her breathing, and her evolving experience of *connection* (to her instrument), the bodily *disconnection* experienced after sustained singing in the artefact, and her eventual loss of strength and the *questioning* of "who" was in control of the soundscape.

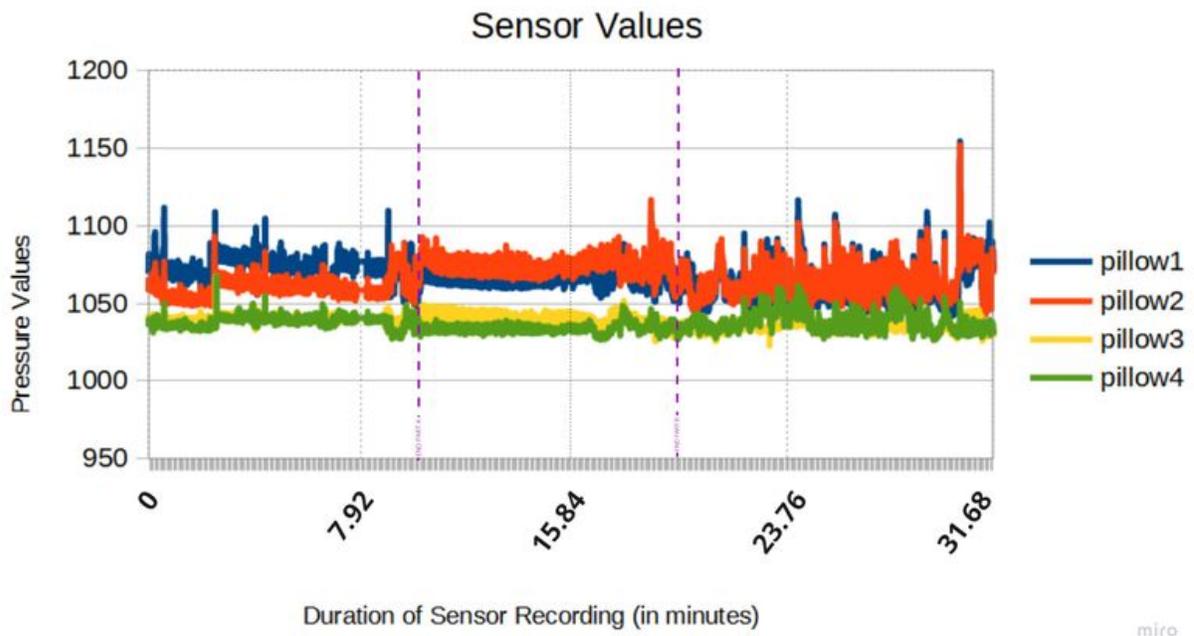

Figure 4. Sample of pillow data visualisation used as notation. Purple segmentation denotes the end of sections.

The encapsulation of Kelsey's experience of singing in the corset through these 3 phases *(connection*, *disconnection* and *questioning*) presents an interesting navigation of a control versus co-performance relationship, facilitated by the experience of singing with and against the interface. The artefact's acceleration of Kelsey's normal physical fatigue levels establishes a transference of "control" of the performance through the *connection* with the object she is interacting with, to a *disconnection* with the object—where control is exchanged—and finally reaching a point at which her body is no longer capable of  producing the range of motion necessary to interact with the machine to "make" it sing (*questioning*). Control is ceded to the artefact, and it transcends its usage as a "thing" to be interacted with - it becomes a sort of independent co-performer through it's impact upon the body and the produced sound, and for it's assumption of control over the soundscape.

Kelsey brought together her experience of singing in the corset and the evolving control relationship it established, coming to 3 states reflective of this experience, drawing further inspiration from concepts identified as central ideas in Whitman's *Leaves of Grass,* namely the corruption of the mind-body-soul.

1. Connection (to-with the body)
2. Disconnection (to-with the body)

3. Questioning (of boundaries between body and artefact)

These became microstructures within the larger composition (see Figure 5), and the sonic and aesthetic aims of these episodes are outlined below.

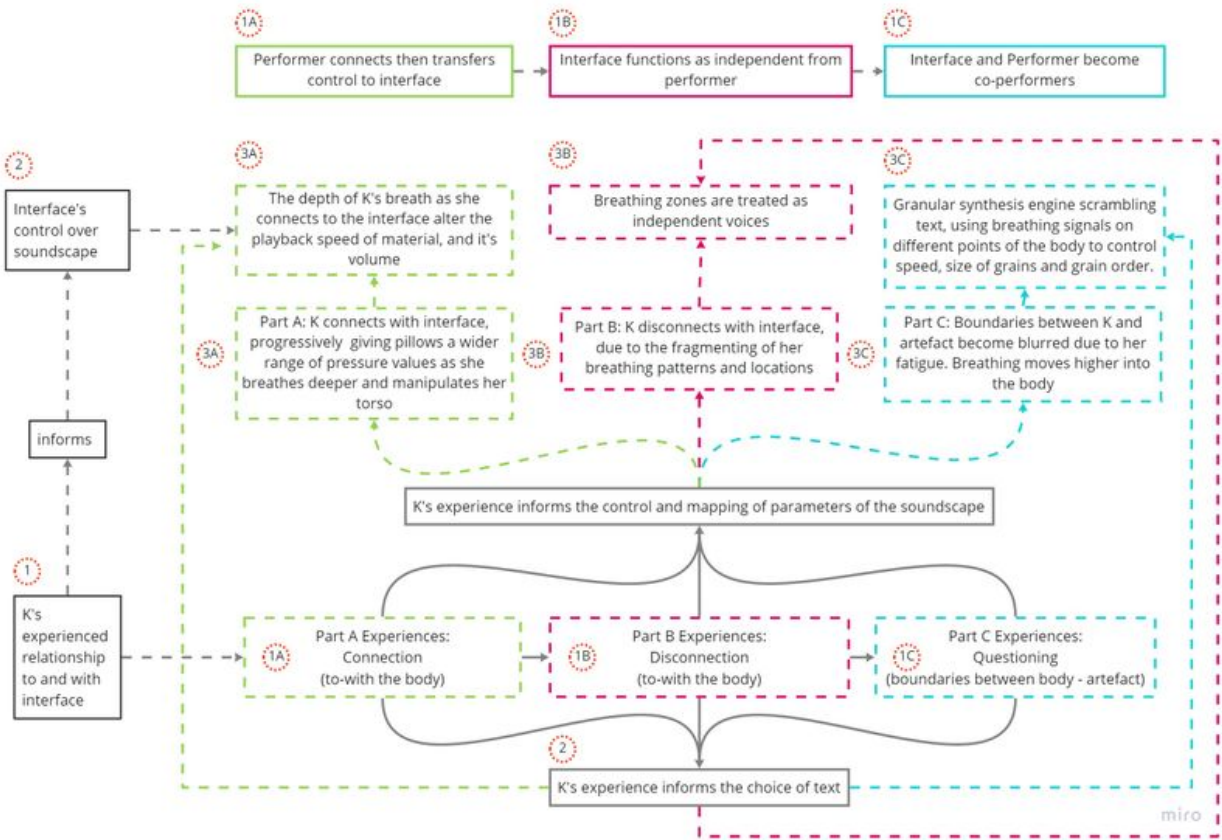

**Figure 5. Outline for impact of experience of singing in interface, and the control vs. co-performer relationship. Used as compositional structure. The structure is labelled numerically to assist reading the table, and to indicate the hierarchy and evolution of control to co-performance relationship.**

## 5.4 The Body Electric performance

### 5.4.1 Connection (and catching breath)

In the first section of the composition (see Figure 5, green), a processing module in MaxMSP that utilised the pillow data to manipulate the playback rate and volume of pre-recorded vocal material was designed. Each pre-recorded line is progressively introduced, and the playback rate and volume is manipulated as Kelsey takes in more air. With each addition of text, the pillows are given more "control" over the rate at which the sonic material is manipulated and distorted, and its volume, when she breathes. The sounds of Kelsey's live breathing—controlling the manipulation of the tape material—is further amplified within the soundscape. They (the artefact and Kelsey) navigate between a control and co-performance relationship in this section, as she connects more with the artefact (by breathing deeper and crushing the pillows), giving more agency in its manipulation of sound by allowing it more 'access' to her breathing. As Kelsey breathes deeper, her movements have greater sonic consequences on the soundscape, and further impact upon her ability to sing with the artefact in later sections, because the corset accelerates her physical fatigue. During this section, Kelsey manipulated her breathing depth and slowly moved her breath around her body to each pillow,  allowing the sonic material to be progressively mutated through her increasing connection to the interface.

## 5.4.2 Disconnection (and dissecting bodies)

In this second section(see Figure 5, pink) Kelsey wanted to separate her breathing from her singing, by exploring the concept of "voiceless singing". The muscles controlling her breathing construct her voice, using the physiological changes in her body that occur normally when she sings, but without her actually making sound. In this portion of the composition Kelsey used her body as a sort of bellows, breathing in different directions, speeds and depths to create a body-motion activated choir with her muscles, using the psychoirtrist~ harmoniser[1] in MaxMSP. She sought to disconnect herself from her breathing in order to use her sensed breathing zones (see Figure 1) as independent musical voices. Each pillow, shaped by and interacting with her changing body, began to "sing" with each other, breaking off into duets and trios, as Kelsey spatialised her breath across each pillow zone (see Figure 1, Panel B). Her body and breath inhabited the role of a co-performer with these pillows, through deliberate *disconnection* of her breath within her body and de-synchronised breathing actions. To further establish the independence of each muscle - each sensor - was mapped in a one-to-many configuration, with each pillow assigned to its own specific "voice" within the psychoirtrist~ harmoniser. This specific harmonizer was uniquely suited to this, as the base patch is configured for 4 control components within each signal transformation. The mapping of each pillow to their own designated control component at each level of the signal transformation, emphasised the de-synchronisation and *disconnection* between her body and breath, and the 4 pillow "voices".

### 5.4.3 Questioning and scrambling bodily boundaries

In this final episode of the work (see Figure 5, blue), biosignal data from the sensors is used to control a granular synthesis engine in MaxMSP which processes the voice in real-time. The functions of the engine, such as grain size, positioning and speed are directly controlled by the pressure values of the 4 sensor pillows. Kelsey used the physical fatigue and disconnection between body and breath that she typically experiences during this sequence to determine the ordering, length and speed of the grains of her voice. The comprehensibility of the text in this episode is determined by the extent to which her fatigued and failing muscles can interact with the corset, pushing the pillows back. Each pillow takes a role in controlling various components of the granular engine, using data streams from pillows placed higher on her body as her breath moves higher as her body begins to "break down".

## 6. Discussion

Through the experience in the design, construction and experimentation of her corset interface in the composition and performance of her work *The Body Electric* ([video via this link](#)) it became apparent that the re-framing of the body as the musical interface facilitated a departure from the typical experience of working with a NIME. Previous work in NIME design has approached the construction of interfaces that harness bodily information such as biosignals or gestural data, working on building a control system  situated within an understanding of the body in space and the creative potential of harnessing electrical signals naturally occurring within the performer's body, or the body's gestural movements. By turning away from this approach, and drawing from principles of soma design, K's utilisation and engagement with her own soma (and the changes to and in it) made the corset not just an interactive interface, but a sort of co-performer which facilitates curious engagements between body, technology and somatic experience.

### 6.1 Interacting *through* and *with* the Body Electric interface

The interactivity offered by this corset also impacted upon Kelsey's relationship *with* and *through* her instrument, causing her to disrupt and adjust her normal vocal practice, specifically her breathing, yielding a range of breathing variations to shape the soundscape. We consider this an example of the novel interactions that emerged through working with the interface. This influenced the approach to composing, where she used the progression of her interaction with the artefact (*Connection - Disconnection - Questioning)* as musical sequences. These were linked to Whitman's notion of corruption, in the changing connection Kelsey experienced with her instrument as a result of the engaging with the interface: *connectionand catching breath* (section 5.5.1) - *disconnectionand dissected bodies* (section 5.5.2) - *questioning and scrambling bodily boundaries* (section 5.5.3).

## 6.2 Control and curious entanglements

Kelsey's interaction with the artefact, and her "control" of the sound-world being transferred to the corset interface throughout performance, was used as a meta-structure for the composition, as was the accelerated 'making and unmaking' of her instrument. The corset provided a lens through which Kelsey's voice is diffracted by the relationship between her and it, and the dissolving of the boundaries between them. This 'controller to co-performer' dynamic, recalls Naccarato's and MacCallum's proposition on subject and object becoming entangled [44]. Kelsey's body and her physiological processes become entangled with, and shaped by, the corset.

Through building a NIME which places the bodily experience of a physiological process as central, it was discovered that an interface designed for the body using soma design principles can open up creative opportunities in examining the nature of the relationship between performer and interface, how both influence each other and the potential this invites for dissolving of the notion of performer-as-controller of object. This presents creative opportunities in re-considering musical interfaces as co-performers in an artistic work, and usage of the experience of wearing and interacting with somaesthetic NIMEs in the development of creative musical works. Our approach to using soma design for developing this NIME, as a design method that puts a strong focus on becoming attentive to and reflecting on first person somatic experiences, shows a great potential for how soma design can enrich the NIME community. Our work contributes with a rich example of a NIME, accompanied by the detailed process behind its design, development and deployment as a musical performance piece. This NIME is grounded in the performer's experience of, and relationship to their bodily changes in performance. Thus, our work extends the work by Martinez-Avila et al [36], offering a concrete example of using soma design for NIMEs, accompanied by reflections on the expanded experience of the Kelsey's work with her interface.

## 7. Conclusion

This paper presents a novel interface for the capture and use of biofeedback signals generated by breathing and subsequent physiological changes caused during singing. It presents a novel approach to NIME design which is grounded in the performer's subjective experience of their body in performance. The creation of this NIME draws from principles of soma design, resulting in an ultra-personalised interface that utilises sensors capable of capturing the subjective physiological changes experienced in the body of the performer. The somatic NIME corset enabled novel interactions and an unexpected "making and unmaking" of the first author's instrument. The harnessing of the biosignal information was further used as a compositional tool in a performative and artistic setting, facilitating the performer's interaction with their body through their body. The findings from this work illustrate the creative potential of designing somatic NIMEs that utilise the bodily experience of a musician, and their experiential relationship and connection to their instrument.

## Footnotes

1. The Max/MSP external *psychoirtrist~* (from Norbert Schnell) transposes and delays a monophonic input multiple times with random variations obtaining a choir effect. ↩

## Citations